\documentstyle[12pt]{article}
\def\rishi{\rangle \! \rangle}

\def\prt{\partial}

\def\d#1{\,{\rm d}#1}

\newcommand{\al}{\alpha'}
\newcommand{\de}{\partial}
\newcommand{\be}{\begin{equation}}
\newcommand{\ba}{\begin{eqnarray}}
\newcommand{\ea}{\end{eqnarray}}
\newcommand{\ee}{\end{equation}}
\newcommand{\db}{\bar{\partial}}
\newcommand{\we}{\wedge}

\newcommand{\lr}{\leftrightarrow}
\newcommand{\f}{\frac}
\newcommand{\s}{\sqrt}
\newcommand{\vp}{\varphi}

\newcommand{\tvp}{\tilde{\varphi}}
\newcommand{\tp}{\tilde{\phi}}
\newcommand{\ti}{\tilde}
\newcommand{\ap}{\alpha}

\newcommand{\ddd}{\cdot\cdot\cdot}
\newcommand{\no}{\nonumber \\}

\begin{document}
\begin{titlepage}
\thispagestyle{empty}
\begin{flushright}
UT-02-23 \\
hep-th/0204234 \\
April, 2002
\end{flushright}

\bigskip

\begin{center}
\noindent{\Large \textbf{
Open Strings in Exactly Solvable Model of \\ \vspace{3mm} 
Curved Spacetime and
PP-Wave Limit}}\\
\vspace{2cm}
\noindent{
Hiromitsu Takayanagi\footnote{hiro@hep-th.phys.s.u-tokyo.ac.jp} }
and \, Tadashi Takayanagi\footnote{takayana@hep-th.phys.s.u-tokyo.ac.jp}
\\
\vskip 2.5em

{\it Department of Physics, Faculty of Science, University of Tokyo\\
Hongo 7-3-1, Bunkyo-ku, Tokyo, 113-0033, Japan}

\vskip 2em

\end{center}

\begin{abstract}
In this paper we study the superstring version of 
the exactly solvable string model constructed by Russo and Tseytlin.
This model represents superstring theory in a curved spacetime and can be
seen as a generalization of the
Melvin background. 
We investigate D-branes in this model as probes of the background geometry
by constructing the boundary states. We find that
spacetime singularities in the model become smooth at high energy
from the viewpoint of open string.
We show that
there always exist bulk (movable) D-branes by the effect of electric flux.
The model also includes Nappi-Witten model as the Penrose limit and 
supersymmetry is enhanced in the limit. We examine this phenomenon 
in the open string spectrum. We also find the similar enhancement
of supersymmetry can be
occurred in several coset models.

\end{abstract}
\end{titlepage}

\newpage

\section{Introduction}
\hspace{5mm}
Recently, many aspects of 
string theory have been uncovered. If we take the perturbative
analysis of conformal field theory as an example, however,
most of the discussions are 
restricted to the curved space not the curved spacetime.
Thus we cannot completely 
answer how to do with the intriguing phenomena peculiar
to curved spacetime such as black holes and spacetime singularities
in the framework of string theory. 
It may seem natural to expect some stringy
resolution of singularities as in orbifold theories. Motivated by this
we would like to study the superstring version
of the bosonic model
 \cite{3RuTs} constructed\footnote{For recent discussions on  
conformal field theoretic approach to time dependent background 
see e.g.\cite{BaHaKeNa,CoCo,Ne,LiMoSe,
ElGiKuRa}.} by Russo and Tseytlin.
 
The best advantage of this model is that it is exactly solvable.
Thus we can compute the closed string spectrum completely. Furthermore,
this background
has spacetime singularities in some parameter regions and thus will be 
suitable
for the study of stringy analysis of spacetime singularities. As particular
limits, this model includes the Melvin background \cite{SM}.
The background is generally non-supersymmetric and includes closed string 
tachyons. One type of tachyons is the localized tachyon discussed recently 
in \cite{closed,Su,TU1,RuTsS,MiYi,DaGuHeMi}. 
There is also another type of the tachyon which appears
when the spacetime has singularities.

The main purpose of this paper is to investigate the geometry of the 
background
by using D-brane probes. Especially, we construct explicit boundary states
in this spacetime. As a result we will show that the geometry becomes 
non-singular at high energy 
from the viewpoint of open string even if there are apparent 
spacetime singularities. Our analysis is parallel with the 
previous studies on D-branes in 
the Melvin background (\cite{DuMo,TU2,TU3}). However, we will find 
several crucial differences from the results in the Melvin model. For
example,
we can always put D0-branes anywhere in this curved spacetime.
Thus we can probe the geometry smoothly, while in the Melvin model we can
construct D0-branes only at the origin for irrational values of parameters
\cite{TU2,TU3}.

This model also includes
 Nappi-Witten model \cite{NaWi} 
as its Penrose limit (pp-wave limit) \cite{Pe} (for recent discussions 
on the duality between string on pp-waves and gauge theory see 
\cite{BeMaNa} and also \cite{ItKlMu,GoOo,ZaSo,pporbifold,
BMNopen,holography,Cv,nsppn,pporbifolde}).
Only in this background there are 
unbroken supersymmetries and thus the supersymmetry
 is enhanced
by taking the Penrose limit of this model. Later we will analyze the similar
enhancement does occur in several coset models. We will also discuss D-branes
in Nappi-Witten model by employing our general results.

This paper is organized as follows. In section 2, we review the model
\cite{3RuTs} and show the detailed analysis of its supersymmetrization.
In section 3, we construct the boundary states of D-branes 
in this model and compute the open string spectra.
In section 4, we consider the enhancement of supersymmetry in the 
Penrose limit of our model as well as the coset models \cite{ZaTs}
and \cite{BrStVl}
from the viewpoint of both closed string
and open string.

\section{Exactly Solvable Superstring Model of \\Curved Spacetime}
\setcounter{equation}{0}
\hspace{5mm}
Here we consider the supersymmetrization of the solvable model \cite{3RuTs}.
This solvable background describes the curved spacetime with four parameters
$R,q_+,\ap,\beta$.
In particular, it includes Nappi-Witten background \cite{NaWi} and the 
Melvin background \cite{SM} as
specific limits. The explicit form of metric $G_{\mu\nu}$, 
NSNS B-field $B_{\mu\nu}$ 
and dilaton $\phi$
of this background is given by
\ba
ds^2&=&-dt^2+dy^2+d\rho^2\no
&+&\f{\rho^2}{1+\ap\beta\rho^2}
\left(d\vp+(q_++\beta)dy-(q_-+\beta)dt\right)
\left(d\vp+(q_+-\ap)dy-(q_-+\ap)dt\right),\no
B_{y\vp}&=&\f{\ap+\beta}{2}\f{\rho^2}{1+\ap\beta\rho^2},\ \ \
B_{t\vp}=\f{\ap-\beta}{2}\f{\rho^2}{1+\ap\beta\rho^2},\no
B_{ty}&=&\left(\f{\ap-\beta}{2}q_++\f{\ap+\beta}{2}q_-+\f{\ap^2+\beta^2}{2}
\right)\f{\rho^2}{1+\ap\beta\rho^2},\ \ \ 
e^{2(\phi-\phi_0)}=\f{1}{1+\ap\beta\rho^2},\label{KK}
\ea
where we introduced the parameters $\ap,\beta,q_+$ and $q_-$. We can 
change the value of $q_-$ by shifting the field $\vp$ such that 
$\vp\to\vp+\lambda t$ and thus $q_-$ is an auxiliary parameter. 
We 
also assume that the coordinate $y$ is compactified and its radius
is denoted by $R$. After the Kaluza-Klein compactification we can obtain
a series of curved Lorentzian backgrounds with electro-magnetic flux
as discussed in \cite{3RuTs}.

If we assume the specific parameter region $\ap\beta<0$, then the 
background becomes singular\footnote{
Obviously, the string coupling $e^{\phi}$ shows the singular behavior.
Furthermore, it is also 
easy to check that the curvature tensors of the metric and B-field diverge 
at the same points. If we take the T-duality, we have also the 
singularity at $1+q_+(q_+ +\beta-\ap)\rho^2=0$.} 
at $\rho_0=1/\s{-\ap\beta}$. However, we 
have the free field representation even in such a case as we will see 
and thus we include these singular cases in our analysis. Later we will 
discuss
how these spacetime singularities affect closed string and open string.

\subsection{World-sheet Sigma Model and Its Free Field Representation}
\hspace{5mm}
The most important advantage of this model is that we can solve the model
exactly by using the free field representation. This is explicitly shown  
in the bosonic theory \cite{3RuTs}. Its extension to supersymmetric 
theory
can be done in a rather straightforward way as we show below
(this procedure is parallel with the Melvin model \cite{SM,TU2} and
the uniform magnetic field model \cite{RT}). 

The background (\ref{KK}) is described by the following sigma-model
\ba
S_1&=&\f{1}{\pi\al}\int 
d\sigma^2 [-\de t \db t+\de y \db y+\de \rho \db \rho\no
&&+\f{\rho^2}{1+\ap\beta\rho^2}\left(\de\vp
+(q_+ +\beta)\de y-(q_- +\beta)\de t\right)
\left(\db \vp+(q_+-\ap)\db y-(q_-+\ap)\db t\right)\no
&&+\f{\al}{4}R^{(2)}(\phi_0-\f{1}{2}
\ln(1+\ap\beta\rho^2))],\label{S1}
\ea
where $R^{(2)}$ is the Ricci scalar of the world sheet.
Here we omitted the fermion terms because they can be simply obtained by
using the superspace formalism as in \cite{TU2}.
The last term of (\ref{S1}) represents the dilaton coupling and below
we will not write this explicitly.

We would like to show how to solve this model exactly by 
examining it in a covariant way without using the light-cone gauge.
First we perform T-duality with respect to $\vp$ (see also \cite{3RuTs,TU2}) 
and obtain
\ba
S_2&=&\f{1}{\pi\al}\int d\sigma^2\Bigl[(\de u+\ap\de\tvp)(\de v+\beta\db\tvp)
+\de\rho\db\rho+\f{1}{\rho^2}\de\vp\db\tvp
+2q_+(\de y\db\tvp-\de \tvp\db y)\no
&&-2q_-(\de t\db\vp-\de \vp\db t )\Bigr], \label{S2}
\ea
where we have defined $u=y-t$ and $v=y+t$. After we replace\footnote{
Here we would like to notice that the time direction $T$ of the free fields
includes the winding number of $\tvp$. This produces an important stringy
correction.} $u,v$ with $U,V$ as follows
\begin{equation}
U=u+\ap\tvp\ (=Y-T) \quad , \quad V=v+\beta\tvp\ (=Y+T),
\end{equation}
we can take the second T-duality 
on $\tvp$,
\ba
S_3&=&\f{1}{\pi\al}\int d\sigma^2
\Bigl[\de U\de V+\de\rho\db\rho+\rho^2(\de\vp'+q_+\de Y-q_-\de T)
(\db\vp'+q_+\db Y-q_-\db T)\Bigr].\no \label{S3}
\ea

If we shift the angular coordinate such that $\vp''=\vp'+q_+ Y-q_- T$, 
then we obtain the free field 
action
\ba
S_3=\f{1}{\pi\al}\int d\sigma^2 [\de U\db V+\de X \db \bar{X}].\label{S4}
\ea
where we defined $X=\rho e^{i\vp''}$ and $\bar{X}=\rho e^{-i\vp''}$. Then
we can see that the fields $U,V,X$ and $\bar{X}$ are all free fields.
In this way we can reduce the original action (\ref{S1}) to the free field
one (\ref{S4}) and thus we can solve the model exactly. 

Next let us examine the mass spectrum. We define the angular momentum
\ba
\hat{J}_L=\f{1}{2\pi i}\oint dz\ j_L(z),\ \ 
\hat{J}_R=-\f{1}{2\pi i}\oint d\bar{z}\ j_R(\bar{z}), \label{J}
\ea
where $j_{L,R}$ is defined as follows
\ba
\label{tvpj}
&&\de \tvp =-\rho^2\de \vp ''+i\psi_{L}\bar{\psi}_{L} \equiv 
i\al j_L,
\no 
&&\db \tvp =\rho^2\db \vp ''-i\psi_{R}\bar{\psi}_{R} \equiv - 
i\al j_R ,
\ea
where we restored the contributions of fermions.
Then we obtain the shifted periodicity of $\tvp$
\ba
\label{tvpperiod}
\tvp(\tau,\sigma+2\pi)=\tvp(\tau,\sigma)-2\pi\al \hat{J}.
\ea

By using the above and the canonical quantization of momenta in the 
original action (\ref{S1}),
the momenta of free fields $Y$ and $T$ are given by
\ba
&&P^{L}_Y+P^{R}_Y=2\left(\f{n}{R}-(q_+ +\f{\beta-\ap}{2})\hat{J}\right),\ \ 
P^{L}_Y-P^{R}_Y=2\left(\f{Rw}{\al}-\f{\ap+\beta}{2}\right),\no 
&&P^{L}_T+P^{R}_T=-2\left(E+(q_-+\f{\ap+\beta}{2})\hat{J}\right), \ \ 
P^{L}_T-P^{R}_T=(\ap-\beta)\hat{J},
\ea
where $n/R,w$ and $E$ represent the Kaluza-Klein momentum, winding number 
in the $y$ direction and the energy. 

The free fields $X,\bar{X}$ and 
$\psi_{L,R},\bar{\psi}_{L,R}$ obey the following 
twisted boundary conditions
\ba
\label{bc}
X(\tau,\sigma+2\pi)&=&e^{2\pi i \gamma}X(\tau,\sigma),\no 
\psi_{L}(\tau,\sigma+2\pi)&=&
e^{2\pi i \gamma}\psi_{L}(\tau,\sigma),\ \ \ 
\psi_{R}(\tau,\sigma+2\pi)=e^{2\pi i\gamma}\psi_{R}(\tau,\sigma),
\label{phase}
\ea
where $\gamma$ is defined by
\ba
\gamma&=&(q_+ +\f{\beta-\ap}{2})wR+\f{\ap+\beta}{2}\al\left(\f{n}{R}-
(q_+ +\f{\beta-\ap}{2})\hat{J} \right)\no
&&+\f{\beta-\ap}{2}\al\left(E+(q_- +\f{\ap+\beta}{2})\hat{J}\right).
\ea

Now we can compute the spectrum of the string model. The spectrum is given 
by $H_{c}=0$, where  $H_c$ is the closed string Hamiltonian
\ba
H_{c}&=&-\f{\al}{2}\left(E+(q_- +\f{\ap+\beta}{2})\hat{J}\right)^2
+\f{R^2}{2\al}\left(w-\f{\al}{R}\f{\ap+\beta}{2}
\hat{J}\right)^2
+\f{\al}{2R^2}\left(n-(q_+ +\f{\beta-\ap}{2})R\hat{J} \right)^2\no
&&+\sum_{i=1}^{6}\f{\al}{2}(p^i)^2-\f{\al}{8}(\ap-\beta)^2
\hat{J}^2+\hat{N_{R}}+\hat{N_L}-\hat{\gamma}(\hat{J_R}-\hat{J_L}),
\label{MS1}
\ea
with the level matching condition
\ba
\hat{N}_R-\hat{N}_L-nw+[\gamma]\hat{J}=0,
\ea
where we define $\gamma \equiv [\gamma]+\hat \gamma$; $[\gamma]$ denotes
the integer part of $\gamma$. Even though this expression 
is the same as in \cite{3RuTs}, we took the
fermions into account (see (\ref{J})) here.
Note that if $\hat \gamma=0$, we must add
the contribution of zero modes of $X$ to $H_c$.

\subsection{Partition Function}
\hspace{5mm}
It is useful to compute the one-loop partition function of 
this string model.
The amplitude can be most easily computed by the path-integral method
in the light-cone gauge as was done in the Melvin model (see 
also \cite{SM,TU1}). We show
the result in Green-Schwarz formulation as follows (for the bosonic model 
see \cite{3RuTs})

\ba
Z(R,q_+,\ap,\beta)&=&(2\pi)^{-7}V_{7}R(\al)^{-5}\int 
\f{(d\tau)^2}{(\tau_2)^6}
\int (dC)^2 \sum_{w,w'\in {\bf Z}}\ 
\f{\theta_{1}(\f{\chi}{2}|\tau)^4\theta_{1}(\f{\ti{\chi}}{2}|\bar{\tau})^4}
{|\eta(\tau)|^{18}
\theta_{1}(\chi|\tau)\theta_{1}(\ti{\chi}|\bar{\tau})}\no
& &\times \exp \Bigl[ -\f{\pi}{\al\tau_2}(
4C\bar{C}-2\bar{C}R(w'-w\tau)+2CR(w'-w\bar{\tau}))\Bigr], \label{PF1}
\ea
where $V_7$ denotes the infinite volume of ${\bf R}^7$ and we have defined 
\ba
\chi=2\beta C+q_+ R(w'-\tau w),\ \ \ti{\chi}=2\ap \bar{C}+q_+ 
R(w'-\bar{\tau} w).
\ea
It is easy to see that the previous spectrum (\ref{MS1}) in
the operator formulation is reproduced by employing the Poisson resummation.
If one uses the Jacobi identity
\ba
\theta_{3}(0|\tau)^3\theta_{3}(\chi|\tau)
-\theta_{2}(0|\tau)^3\theta_{2}(\chi|\tau)-
\theta_{4}(0|\tau)^3\theta_{4}(\chi|\tau)=2\theta_{1}(\f{\chi}{2}|\tau)^4
\label{JA1},
\ea
then this explicitly represents the path-integral in the NS-R formulation 
with correct
type II GSO-projection.

Next we would like to mention the T-duality symmetry. The result is the same
as in the bosonic model \cite{3RuTs}. Thus we obtain the following results.
\ba
Z(R,\ap,\beta,q_+)=Z(\al/R,q_+,\beta-\ap+q_+,\ap)
=Z(\al/R,\ap-\beta-q_+,-q_+,-\beta).
\ea
In other words, the model is invariant under the exchange of
parameters
\ba
R\lr\al/R,\ \ \f{\ap+\beta}{2}\lr
\ti{q}=\f{2q_++\beta-\ap}{2},\ \ \ \ \ (\beta-\ap=\mbox{fixed})
\label{tdual}.
\ea
We can also see that the superstring theory in the Melvin 
background \cite{SM} is 
included as a particular case $\beta=\ap,\ q_-=-\beta,\ q_+=q$. 
In particular 
this shows that this background includes the two dimensional 
oribfolds ${\bf{C}}/{\bf{Z}}_N$ in type II and type 0 string theory 
\cite{TU1}. 

For generic values of parameters as we can see from (\ref{PF1}),
the partition function does not
vanish. This means that this model is non-supersymmetric in general.
\subsection{Penrose Limit and Nappi-Witten Model}
\label{sec:PPNW}
\hspace{5mm}
 We can also see that the 
sigma model (\ref{S1}) is equivalent to
the (compactified) Nappi-Witten model \cite{NaWi} (see also \cite{nsppo,
nsppn} and references their in)
\ba
S&=&\f{1}{\pi\al}\int 
d\sigma^2 [\de u\db v+\beta\rho^2\de u\db\phi+\rho^2\de\phi\db\phi],
\label{NP}
\ea
if we set $\ap=q_{\pm}=0$ (or $\beta=q_{\pm}=0$) as shown in
\cite{3RuTs}. 
The spectrum of this model is given by (see also \cite{RuTs})
\ba
\f{\al}{2}E^2+\al\beta E\hat{J}_{R}
=\f{R^2}{2\al}w^2+\f{\al}{2R^2}n^2+\beta wR\hat{J}_R-
\al\beta\f{n}{R}\hat{J}_R+\hat{N}_L+\hat{N}_R,
\ea
where we assumed $0\leq \gamma < 1$.
Note that this spectrum is invariant under the simple 
T-duality transformation
$R\to \al/R$.
We can also compute its partition function\footnote{We are
grateful to Y.Sugawara for pointing our the mistakes
in this equation of the previous version.}
 as a limit of (\ref{PF1})
\ba
Z(R,\beta)\!=\!(\al)^{-\f92}V_{7}
\ti{V}\!\int \f{d\tau^2}{\tau_2^{5}}\!\!\sum_{w,w'\in {\bf Z}}\!\!
e^{-\f{\pi R^2}{\al\tau_2}|w'-w\tau|^2}\!\!\!\!\!\!
\f{\theta_1(\beta R(w'-\tau w)/2|\tau)^4\ \theta_1(0|\bar{\tau})^4}
{(w'-\tau w)\ \eta(\tau)^{9}
\eta(\bar{\tau})^{12}\ \theta_1(\beta R(w'-\bar{\tau} w)|\tau)},\
\ \ \label{pn}
\ea
where $\ti{V}$ denotes a divergent factor coming from the extra zero-modes
along $X,\bar{X}$ direction. Note that this partition function does 
vanish\footnote{If
we take the non-compact limit $R=\infty$ (or equally $R=0$ by T-duality), 
then the partition function
is equivalent to that in flat space \cite{nsppo}. However this does not 
imply the 
equivalence between both models since the interactions are obviously 
different. Later we will find the open string spectrum in the 
Nappi-Witten model which differs from that in flat space.}
because $\theta_1(0|\tau)=0$
and this is consistent with the fact that Nappi-Witten background preserves
partial supersymmetries \cite{nsppo,RuTs}. We can also see that 
this is the only case of
vanishing partition function. Note that
this special background satisfies the condition $B_{y\vp}=B_{t\vp}$.
It would also be useful to see that the partition function (\ref{pn})
shows that the background can be regarded as an orbifold with respect to
only left-moving sectors
if we assume the fractional case $\beta R=\f{k}{N}\in {\bf Q}$.

These facts can be understood by taking the Penrose limit \cite{Pe,BF}
of (\ref{KK})
\ba
\ti{u}=u,\ \ \Omega^2\ti{v}=v,\ \ \Omega\ti{\rho}=\rho\ 
\mbox{and}\ \Omega\to 0,
\ea
which leads to the following (rescaled) metric
\ba
(ds')^2=\Omega^{-2}(ds)^2=d\ti{u}d\ti{v}+d\ti{\rho}^2
+\ti{\rho}^2(d\vp+(q_1+\beta)d\ti{u})
(d\vp+q_1d\ti{u}),
\ea
where we defined $q_{\pm}=q_1\pm q_2$. 
Thus we come back to the specific
model $\ap=q_2=0$. Since the Penrose limit decompactifies the circle in the 
$y$ direction, 
we can set $q_1=0$ by the redefinition of coordinates and we finally obtain 
the Nappi-Witten model. Notice that after we take the Penrose limit of the
non-supersymmetric model, we eventually find the supersymmetric background.
Later we will also 
give some other examples of similar supersymmetry enhancement
in the Penrose limit.

\subsection{Closed String Tachyons and Spacetime Singularity}
\hspace{5mm}
In general the background (\ref{KK}) is non-supersymmetric as can be 
seen from the fact that 
 the partition function of this background 
does not vanish. Thus it is very interesting to discuss the instability
of the model. In the bosonic string model some relevant discussions were 
given in \cite{3RuTs}. Here we would like to consider the tachyonic 
instability in the superstring model. We identify a tachyonic mode with
an excitation which has a non-zero imaginary part\footnote{ 
Since in this case $Im E\neq 0$ the value of $\gamma$ also becomes imaginary,
the twisted boundary condition (\ref{phase}) may be related to the Lorentzian
orbifold discussed in \cite{HoSt,BaHaKeNa,CoCo,Ne,LiMoSe}.}
of energy $E$ in the 
spectrum
(\ref{MS1}).

In this model tachyons arises due to two reasons. The first reason is the 
presence of the term which is proportional to $\hat{\gamma}$. The effect
of this is the existence of
localized tachyons\footnote{For rational values of parameters 
\cite{TU1} the existence of bulk tachyons (as in type 0 theory) is also
possible.} \cite{closed,Su,RuTsS,MiYi,DaGuHeMi} and 
they have already appeared
in the Melvin background \cite{SM} $\ap=\beta$
(for
discussions of closed string tachyon condensation
in Melvin background see e.g.\cite{CG,GS,RM,Su,TU1,RuTsS,MiYi,DaGuHeMi}).
This type of tachyonic modes has non-zero values of $n$ or $w$. 

The second reason is that there is the term which is proportional 
to $-\hat{J}^2$ in the spectrum (\ref{MS1}). This leads to tachyons even if
we set $n=w=0$. Let us assume $\ti{q}\equiv q_+ +\f{\beta-\ap}{2}=0$ for 
simplicity and
consider gravitons ($\hat{N}_L=\hat{N}_R=0$) 
which has the spin $\hat{J}_{L}=\hat{J}_{R}=\pm 1/2$. Then we can see that
the mode is tachyonic if and only if $\ap\beta<0$. As in the bosonic case 
\cite{3RuTs} we can speculate that this instability occurs
due to the spacetime singularity of the background (\ref{KK}) at
the fixed $\rho_0$ such that $1+\ap\beta\rho_0^2=0$. Notice that 
this tachyon field is not localized one since the corresponding 
off-shell (tachyonic) vertex operator has the vanishing value of 
$E+(q_-+\f{\ap+\beta}{2})\hat{J}$.
We can also show that 
such tachyons do not exist in the RR-fields. 

One of the most interesting questions is what background this unstable model
will decay into. If we remember the arguments that the Melvin background
will become the ordinary type II string theory \cite{GS,DaGuHeMi}
after the closed string tachyon condensation, then it seems to be natural to
identify the decay product of the general non-supersymmetric
background $\ap\beta>0$ with the Nappi-Witten
background. In the case of $\ap\beta<0$ (singular spacetime) 
we cannot even speculate the answer
because we must consider the condensation of the `new type' (not localized) 
tachyons. We would like to leave these as future problems. Instead we will 
later
 discuss whether we can probe this spacetime singularity by using D-branes.

\subsection{Higher Dimensional Generalizations}
\hspace{5mm}
It is also possible to construct higher dimensional models. For example,
we can generalize the action in the Green-Schwarz formalism as in the case
of Melvin background \cite{TU1,RuTsS} such that there are $n\ge 2$ angular
coordinates $\vp_i\ \ (i=1,\ddd,n)$ 
by introducing $3n$ parameters $\ap_i,\beta_i$ and $q_{+i}$. Then the 
partition function $(n=4\ \mbox{case})$ 
is given by replacing the theta-function part of (\ref{PF1}) with
\ba
\sum_{w,w'\in {\bf Z}}
\f{\left(\sum_{a=2}^{4}(-1)^{a+1}\prod_{i=1}^4\theta_a(\chi_i|\tau)\right)
\left(\sum_{b=2}^4(-1)^{b+1}\prod_{j=1}^4\theta_b(\ti{\chi_j}|\tau)\right)}
{\prod_{i=1}^4\theta_1(\chi_i|\tau)\prod_{i=1}^4\theta_1(\ti{\chi_i}|\tau)}.
\ea

We can see that there exist supersymmetric
models which have the vanishing vacuum amplitude 
(e.g. $\sum_{i}(\pm\chi_i)=0$)
in the same way as in \cite{TU1,RuTsS}.

\section{Boundary States in Curved Spacetime}
\hspace{5mm}
Here we consider D-branes in the solvable
model (\ref{S1}) by using the boundary state formalism. 
This will give the generalization of the arguments on D-branes in 
two parameter 
Melvin
background \cite{TU2} ($\alpha = \beta$). See also \cite{DuMo} for an
independent analysis of D-branes in the
one parameter (Kaluza-Klein Melvin) model.

Here we consider the general case $\alpha \neq \beta$ and a D-brane
in this model can be regarded as a probe in the curved spacetime.
Finally we will find the following two novel facts.

To begin with, there is the term $-\frac{(\alpha -\beta)^2}{8}
\alpha^{\prime}\hat J^2$  in the closed string Hamiltonian
(\ref{MS1}) and this changes the open string Hamiltonian.
In the case of D0-branes in the free field representation,
for instance, we will see that this effect
leads to the term proportional
to $\sin^2 [\frac{\pi(\alpha-\beta)}{2}\alpha^{\prime}E]$ in
the open string Hamiltonian. Though we can interpret this in terms 
of the curved metric in the low energy limit,
this includes a stringy correction in the high energy region. 
We will see that this stringy correction makes the open string 
spectrum smooth and thus we cannot see the spacetime
singularity in the high energy region.

The second result is that we can always move D0-branes by performing 
the Lorentz boost in the direction $Y$ away from the origin
when $\alpha \neq \beta$. This contrasts strikingly with the Melvin model
($\alpha = \beta$). The latter does not allow 
movable D0-branes (bulk D0-branes) if the value of 
$\beta\al/R$ is irrational \cite{TU2}.
This means that there exist the bulk
D-branes for any parameters if $\alpha \neq \beta$. 
In the original coordinate picture,
this boost corresponds to adding the electric flux. And then this 
flux stabilizes the bulk D-branes.

\subsection{Boundary Conditions}
\hspace{5mm}
Let us find the consistent boundary conditions. 
If we consider them in the free field representation $(T,Y,X,\bar{X})$, 
Neumann and Dirichlet boundary condition are obviously allowed
(later we will discuss D-branes in the original coordinates to
investigate their geometrical properties.)

The boundary conditions in the $T$ and $Y$ direction are given by
\begin{equation}
T:\left\{\begin{array}{rll}
\mbox{Neumann :} \quad \partial_{\tau}T|_{\tau=0}|B \rangle=0 
\quad &\rightarrow \quad \displaystyle
\Big[2E +(2q_- +\alpha +\beta)\hat J\Big]|B \rangle=0 ,\\
\mbox{Dirichlet :}\quad \partial_{\sigma} T|_{\tau=0}|B \rangle=0 
\quad &\rightarrow \quad \displaystyle (\beta -\alpha)\hat J|B \rangle=0 ,
\end{array}\right.
\label{boundary condition in T frame}
\end{equation}
\begin{equation}
Y:\left\{\begin{array}{rll} 
\mbox{Neumann :} &\quad \partial_{\tau}Y|_{\tau=0}|B \rangle=0 
\quad &\rightarrow \quad \displaystyle
\Big[\frac{2n}{R} - (2q_+ +\beta -\alpha)\hat J\Big]|B \rangle=0 ,\\
\mbox{Dirichlet :} &\quad\partial_{\sigma} Y|_{\tau=0}|B\rangle=0 
\quad &\rightarrow \quad \displaystyle
\Big[-\frac{2Rw}{\alpha^{\prime}}+(\alpha+\beta)\hat J\Big]|B\rangle=0,
\end{array}\right. \label{boundary condition in Y frame}
\end{equation}
where the coordinates $(\sigma,\tau)$ represent the 
world-sheet of the closed string channel.
In the most part of this paper, we consider D-branes which satisfy the
Dirichlet-Dirichlet boundary condition
in the $(X,\bar X)$ directions since they can be used as probes of the 
background. The D-branes of Neumann-Neumann boundary conditions can also be
examined similarly\footnote{Though Neumann-Dirichlet 
boundary condition can also be allowed, we will not discuss 
this case in the present paper.}.
In these cases, we can verify 
\begin{equation}
\hat J|B \rangle = \hat J_0|B \rangle,
\end{equation}
where $|B \rangle$ is the boundary state and $\hat J_0$ is the bosonic
zero mode contribution to the angular momentum $\hat J$.

Since we are
interested in D-branes which are 
not localized in the time like direction, below we mainly assume 
that the 
field $T$ satisfies Neumann boundary condition. 
In the last of this section
we will briefly mention D-brane instantons.
Thus we would like to show the detailed analysis of D0-branes and D1-branes 
whose 
boundary conditions are 
$(T,Y,X)=(N,D,D)$ and $(T,Y,X)=(N,N,D)$, where ``$N$'' means
Neumann and ``$D$'' Dirichlet. The D2-branes $(T,Y,X)=(N,D,N)$ and 
D3-branes  $(T,Y,X)=(N,N,N)$ can be treated almost in the same way as
D0 and D1-branes. 

\subsection{Bulk D-branes}
\hspace{5mm}
Now we consider the boundary conformal field theory of 
bulk D0 and D1-branes in the free field representation 
 of boundary
states. The term `bulk' means that the brane can leave from the
origin $\rho=0$ and thus it can probe the geometry of the spacetime. 
There are also D-branes which are fixed at the origin.
We call the latter `fractional' as in the orbifold theory \cite{DoMo,DiDoGo}.
Generally, a bulk D-brane is consist of finite numbers of 
fractional D-branes.
We show the detailed analysis only for bulk D0-branes since
bulk D1-branes can be obtained by taking the 
T-duality (\ref{tdual}) in the $Y$ 
direction.
The fractional D-branes are also discussed in the appendix \ref{Ap:frac}.
\subsubsection*{Bulk D0-brane}
\label{sec:bulk D0 at non zero rho}
\hspace{5mm}
The boundary conditions $(T,Y)=(N,D)$ require that
the zero modes should satisfy
\begin{equation}
E +\Big(q_- +\frac{\alpha +\beta}{2}\Big)\hat J=0 \quad , \quad
Rw-\frac{(\alpha+\beta)}{2} \alpha^{\prime} \hat J=0.
\label{TNYD gluing condition}
\end{equation}
By using these conditions, 
we can write the closed string Hamiltonian (see (\ref{MS1})) 
which acts on $|B \rangle$ as
\begin{equation}
2H_c =\frac{\alpha^{\prime}}{2}(p^i)^2+
\frac{\alpha^{\prime}}{2R^2}\Big[
n-\Big(q_+ + \frac{\beta -\alpha}{2}
\Big)R\hat J_0 \Big]^2 -\frac{(\alpha - \beta)^2}{8}\alpha^{\prime}\hat
J_0^2
+\hat N_L + \hat N_R +\hat \gamma
(\hat J_R -\hat J_L),
\label{TNYD Hamiltonian}
\end{equation}
where $\gamma$ is
\begin{equation}
\gamma = \frac{\alpha^{\prime}}{2}(\alpha + \beta)\frac{n}{R}.
\label{gammma TNYD}
\end{equation}
Notice that $\gamma$ depends only on $n$, not on $\hat J_0$. 

When $\frac{\alpha^{\prime}(\alpha + \beta)}{2R}$ is rational ($\equiv 
\frac{k}{N}\in{\bf Q}$)\footnote{We assume that $k$ and $N$ are coprime.},
there are bulk D0-branes as in the case of the Melvin 
background\footnote{As we 
can see from eq.(\ref{gammma TNYD}),
we have only to change $\beta$ to $\frac{\alpha +\beta}{2}$.}.
We can find that a bulk D0-brane at $\rho=0$ consists of $N$ fractional
D0-branes whose positions in the $Y$ direction are given by
$y_0,y_0+\frac{2\pi R}{N},\cdots$ (see appendix \ref{Ap:frac}).

Then let us move the bulk
D0-brane from $\rho =0$. In the case of $\rho \neq 0$,
the states of $\hat J_0 \neq 0$ are allowed.
First, from eq.(\ref{TNYD gluing condition}) we obtain the constraint
\begin{equation}
\qquad
\hat J_0 \in N{\bf Z},\label{TNYD boundary condition again}
\end{equation}
and thus we find that the bulk D0-brane at $\rho \neq 0$ should consist of
$N$ fractional D0-branes located at $N$ points
$X=\rho,\ \rho e^{\f{2\pi i}{N}},\ddd$, in the $X$
direction such that the bulk D0-brane is invariant under the action 
$X\rightarrow e^{\frac{2\pi i}{N}}X$.
We can also understand\footnote{
The world-volume theory of such bulk D-branes seems to be described
by a sort of quiver gauge theory with monodromy discussed 
in \cite{OkSu,MiYi}. We thank Y.Sugawara for 
pointing out this to us.}this fact from the explicit form of
the boundary state (appendix \ref{Ap:Be}). 

Now let us see whether the bulk D0-brane can exist at $\rho \neq 0$ in
this case. In order to check this we must examine
the Cardy's condition \cite{Ca}. 
The vacuum amplitude (assuming
$k$=even for simplicity)
\begin{equation}
{\cal A} = \int \d s \langle B|\frac{\alpha^{\prime}}{2}e^{-2sH_c}|B\rangle,
\end{equation}
is given by 
\begin{equation}
{\cal A}_{bulk}=\frac{\alpha^{\prime}V_0}{8\pi R}
\left(\frac{NT_0}{2}\right)^2\int \d s \, (2\pi \alpha^{\prime}s)^{-3}
\eta(\tau)^{-12}[\theta_3(0|\tau)^4-\theta_4(0|\tau)^4
-\theta_2(0|\tau)^4]\ Z_{0}(\tau),\label{vacuum}
\end{equation}
where $V_0$ denotes the volume in the time direction and we defined
$T_0 = \sqrt{\pi}(2\pi \sqrt{\alpha^{\prime}})^3$. We have also written 
the zero-mode part
of the amplitude as
$Z_{0}(\tau),\ \tau\equiv\f{is}{\pi}$ and it is given by
\ba
Z_{0}=\langle B_{0}|e^{-s[\al \vec{k}^2/2-\f{(\ap-\beta)^2}{8}\al \hat J_0^2+
\f{\al}{2R^2}(n-\ti{q}R\hat J_0)^2]}|B_{0}\rangle, \label{am}
\ea
where we defined $\ti{q}=q_+ +\f{\beta-\ap}{2}$ (see (\ref{tdual}))
and $\vec{k}$ is the momentum in the $X$ direction. Notice that
we can find that the above vacuum amplitude (\ref{vacuum}) does 
vanish and Bose-Fermi degeneracy occurs.
Below we first assume $\ap+\beta=0$ ($k=0,N=1$) and later
consider general cases. Then we can write the zero-mode part of
the boundary state as
\ba
|B_{0}\rangle =\sum_{n}\frac{1}{(2\pi)^2}
\int (d\vec{k})^2\ e^{i\vec{k}\cdot\vec{x}}|\vec{k}\rangle \otimes 
|n\rangle,
\ea
where $\vec{x}=(\rho\cos \phi,\rho\sin \phi)$ is the position
of the D0-brane in the $X$ direction and the $y$-momentum eigenstate 
$|n \rangle$ is 
normalized as $\langle n|n^{\prime}\rangle =\delta_{n n^{\prime}}$.
Next we perform the modular transformation ($s=\pi/t$)
of this amplitude. After the 
the Poisson resummation of 
$n$ and the Gauss integration of the open string
energy $E$, we obtain
\ba
Z_{0}&=& \sum_{w}
\s{\frac{2R^2t}{\alpha^{\prime}}}\s{\frac{2\pi\alpha^{\prime}}{s}}
\int (dE)\ \langle B^{\prime}_{0}|\ e^{\f{2\pi^2\al}{s}E^2
-\f{s\al \vec k^2}{2}-2\pi t\f{R^2w^2}{\al}+i\pi
\left((\ap-\beta)\al E+2\ti{q}Rw\right) \hat J_0}|B^{\prime}_{0}\rangle \no
&=& \sum_{w}\frac{2Rt}{(2\pi)^2}\int (d\vec{k})^2 (dE) e^{\f{2\pi^2\al}{s}E^2
-\f{s\al \vec k^2}{2}-2\pi t\f{R^2w^2}{\al}}e^{i\vec{k}\vec{\Delta x}},
\ea
where we defined $|B^{\prime}_0 \rangle $ as the zero mode 
part in the $X$ direction.
Note that we have employed the fact that
 the operator $e^{i\theta \hat J_0}$ acts as the rotation
$k_1+ik_2\to e^{i\theta}(k_1+ik_2)$ (or equally
$x_1+ix_2\to e^{i\theta}(x_1+ix_2)$). The length of $\vec{\Delta x}$
is given by
\ba
|\vec{\Delta x}|=
\left|2\rho \sin\left(\f{\pi(\ap-\beta)}{2}\al E
+\pi\ti{q}Rw\right)\right|.
\ea

Then after we integrate out $\vec k$,
the amplitude becomes
\begin{equation}
Z_{0} = \sum_{w}
\frac{2Rt^2}{\pi \alpha^{\prime}}\frac{1}{2\pi}\int \d E e^{-2\pi tH_o},
\label{non trivial Z in N=1}
\end{equation}
where $H_o$ is given by
\ba
H_o=-\al E^2+\f{R^2w^2}{\al}+\f{\rho^2}{\pi^2\al}
\sin^2\left(\f{\pi(\ap-\beta)}{2}\al E
+\pi\ti{q}Rw\right).
\label{non trivial H open in N=1}
\ea

In order to generalize this result for $\f{(\ap+\beta)\al}{2R}=k/N$,
we have only to change eq.(\ref{non trivial Z in N=1}) and
(\ref{non trivial H open in N=1}) about a few points (see 
appendix \ref{Ap:Be} for detail). 
Finally, we obtain
the open string 1-loop amplitude
\begin{equation}
{A}_{bulk}= 2\times \int_0^{\infty}\frac{\d t}{2t}\mbox{Tr}_{NS-R}
\Big[\frac{1+(-1)^F}{2}e^{-2\pi tH_{o}} \Big],
\label{1-loop amplitude nonzero rho}
\end{equation}
where the open string Hamiltonian $H_o$ is written as
\begin{equation}
H_o = -\alpha^{\prime}E^2 +\frac{R^2w^2}{\alpha^{\prime}N^2}
+ \frac{\rho^2}{\pi^2\alpha^{\prime}}
\sin^2 \Big[ \frac{(\alpha -\beta)}{2}\pi \alpha^{\prime}E+
\pi \tilde q R\frac{w}{N} + \frac{\pi m}{N}\Big]
 + \hat N.
\label{open Hamiltonian at non zero rho}
\end{equation}
The trace $\mbox{Tr}_{NS-R}\equiv\mbox{Tr}_{NS}-\mbox{Tr}_R$ means
\begin{equation}
\mbox{Tr}_{NS-R}= \frac{V_0}{2\pi}\int \d E \, \sum_{m=0}^{N-1} 
\sum_{w=-\infty}^{\infty}\, \cdots .\label{trace}
\end{equation}
We can easily 
see from eq.(\ref{1-loop amplitude nonzero rho}) and (\ref{trace}) 
that the result satisfies the Cardy's condition\footnote{If 
we consider the limit $\rho \rightarrow 0$,
$$
H_o \rightarrow -\alpha^{\prime}E^2
+\frac{R^2w^2}{\alpha^{\prime}N^2} + \hat N,
\quad \mbox{Tr}_{NS-R} \rightarrow \frac{NV_0}{2\pi}\int \d E 
\sum_{w=-\infty}^{\infty}\, \cdots . 
$$
The above calculations are consistent with the results at $\rho =0$
(\ref{ZO bulk D0}) and (\ref{open Hamiltonian at origin}) analyzed in the
 different way.}.

\subsubsection*{Bulk D1-brane}
\hspace{5mm}
By using T-duality (\ref{tdual}) in the $Y$ direction, we can obtain 
the result of 
bulk D1-brane.
The open string Hamiltonian is given by
\begin{equation}
H_o = -\alpha^{\prime}E^2 +\alpha^{\prime}\frac{n^2}{R^2N^2}
+ \frac{\rho^2}{\pi^2\alpha^{\prime}}
\sin^2 \Big[ \frac{(\alpha -\beta)}{2}\pi \alpha^{\prime}E+
\pi  \frac{(\alpha +\beta)}{2R}\alpha^{\prime}\frac{n}{N} + 
\frac{\pi m}{N}\Big]
 + \hat N,
\label{open Hamiltonian of D1}
\end{equation}
where we defined $(q_+ +\frac{\beta -\alpha}{2})R=\frac{k}{N}$.

\subsection{D-branes as Probes}
\hspace{5mm}
Here we will further study a bulk D$p$-brane ($p$=0,1) in this model 
so that we can probe the geometry for the fixed values of $\rho$.

Let us first consider the bulk D1-brane and assume
$\ti{q}=0$ for simplicity. If we consider the 
low energy limit (or equally $\al\to 0$ limit)
\begin{equation}
E\ll \f{1}{(\ap-\beta)\al}\, , \quad p_y = \frac{n}{R}\ll 
\f{1}{(\ap+\beta)\al} 
\, 
\end{equation}
then we can expand eq.(\ref{open Hamiltonian of D1}) as
\begin{equation}
H_o = -\alpha^{\prime}\Big[1-\frac{(\alpha -\beta)^2}{4}\rho^2\Big]E^2
+\alpha^{\prime}\frac{(\alpha^2 -\beta^2)}{2}\rho^2 Ep_y
+\alpha^{\prime}\Big[1+\frac{(\alpha +\beta)^2}{4}\rho^2\Big]p_y^2+\ddd 
\label{low}.
\end{equation}
This result can be interpreted as the expression $H_o = \al G^{\mu\nu}
p_{\mu}p_{\nu}+\ddd$ and is indeed consistent with the original metric 
(see (\ref{KK}))
\begin{equation}
G^{tt}= -1+\frac{(\alpha -\beta)^2}{4}\rho^2 \, , \quad
G^{ty}= \frac{(\alpha^2-\beta^2)}{4}\rho^2 \, , \quad
G^{yy}= 1+ \frac{(\alpha +\beta)^2}{4}\rho^2.
\end{equation}

On the other hand, in the high energy region we have a non-trivial $\al$
correction as can be seen from the sin factor in 
(\ref{open Hamiltonian of D1}), which comes from the winding number
of time direction. 
 This leads to 
the intriguing fact that there always exists a non-tachyonic pole\footnote{
Here we should note that tachyonic poles ($Im E\neq0$) can also exist
in the equation $H_{o}=0$. 
However, we cannot answer whether these poles are physically relevant.}
($Im E=0$) as the solution of $H_{o}=0$ for any open string modes
 in spite of the 
presence of spacetime singularities and closed string tachyons.
 Note that if the low energy 
expression (\ref{low}) held for any $E$, 
the spectrum would include only
open string tachyons if $1+\ap\beta\rho^2<0$, which just corresponds 
to the existence 
of spacetime singularity as discussed in section 2.4.
We can also see that if we assume the large value of $\rho$,
there can be many (non-tachyonic) poles in the low 
energy region, while the behavior at high energy 
approaches that in the flat space.

Any way we have observed that the stringy effect seems to 
make the spacetime singularity smooth at high energy 
from the viewpoint of open string.

A similar result can be obtained for the open strings on D0-branes and 
the low energy limit of the spectrum is given as follows
\ba
H_o=-\al E^2(1-\f{(\ap-\beta)^2}{4}\rho^2)+\ddd .
\ea
Again the spacetime singularity cannot
be seen in the high energy region. 

\subsection{Boosted D-branes}
\hspace{5mm}
It is also interesting to consider a bulk D0-brane 
which is boosted in the $Y$ direction since we are discussing the curved 
spacetime (for a review of boundary states including boosted branes 
see \cite{di}). The boundary condition is given by
\ba
&\quad \partial_{\tau}(T+vY)|_{\tau=0}|B \rangle =0  :& \quad
2E+(2q_- +\alpha + \beta)\hat J_0 = v \Big[
\frac{2n}{R}-(2q_+ +\beta -\alpha)\hat J_0 \Big],\no
&\quad \partial_{\sigma}(vT+Y)|_{\tau=0}|B\rangle=0  :& \quad
v(\alpha -\beta)\hat J_0 = -\frac{2Rw}{\alpha^{\prime}}+
(\alpha +\beta)\hat J_0,
\label{boosted D0 boundary condition}
\ea
where $v$ is the velocity of D0-brane.
Then the closed string Hamiltonian 
changes into
\ba
&&2H_c = \frac{\alpha^{\prime}}{2}(p^i)^2+(1-v^2)
\frac{\alpha^{\prime}}{2R^2}\Big[n-\Big(q_+ +\frac{\beta -\alpha}{2}
\Big)R\hat J_0\Big]^2
-(1-v^2)\frac{(\alpha -\beta)^2}{8}\alpha^{\prime}\hat J_0^2
\no &&\qquad+\hat N_L +\hat N_R -\hat \gamma (\hat J_R - \hat J_L),
\ea
where $\gamma$ is given by
\begin{equation}
\gamma
=\frac{\alpha^{\prime}}{2R}
\Big[\alpha+\beta +v(\beta -\alpha) \Big]n \equiv \gamma_{v0}n.
\label{boosted gamma}
\end{equation}
Thus we find that in the case of $\ap\neq \beta$, we can 
always make the value of $\gamma_{v0}$ rational by
tuning the velocity,
even if $\frac{(\alpha +\beta)\alpha^{\prime}}
{2R}$ is irrational. This means that bulk D0-branes always
exist for any value of $\frac{(\alpha +\beta)\alpha^{\prime}}{2R}$. This is 
crucially different from the result in Melvin model ($\beta =\alpha$),
where a bulk D0-brane does not exist when $\beta\al/R$ is irrational.

Now let us check that boosted bulk D0-branes satisfy the Cardy's 
condition\footnote{We show that boosted fractional D0-branes satisfy 
the Cardy's condition in appendix \ref{Ap:frac}.}. First, when $\gamma_{v0}=
\frac{k}{N}$ and $k=$even, the vacuum amplitude (\ref{vacuum}) changes to
\begin{equation}
{\cal A}_{bulk}=\sqrt{1-v^2} \times \frac{\alpha^{\prime}V_0}{8\pi R}
\left(\frac{NT_0}{2}\right)^2\int \d s \, Z_{0}(v,s)\cdots , \label{vacuum v}
\end{equation}
where the factor $\sqrt{1-v^2}$ is due to the Lorentz contraction.
But this factor is absorbed\footnote{We set $N=1$ in eq.(\ref{absorb})
for simplicity. For general values of 
$N$, $v$-dependence is the same as in $N=1$.}
 when we use the Poisson resummation formula 
\begin{equation}
\sqrt{1-v^2}\sum_{n}
\exp \Big[-\frac{\alpha^{\prime}(1-v^2)}{2R^2t}(n-\tilde qR\hat J_0)^2\Big]=
\sqrt{\frac{2R^2t}{\alpha^{\prime}}}
\sum_{w}\exp \Big[-\frac{2\pi R^2t}{\alpha^{\prime}(1-v^2)}
w^2+2\pi i \tilde q R w \hat J_0\Big].\label{absorb}
\end{equation}
Then, we can find from eq.(\ref{absorb}) that the boosted bulk 
D0-branes also
satisfy the Cardy's condition. Finally, the open string 
Hamiltonian changes to
\begin{equation}
H_o = -\alpha^{\prime}E^2 +\frac{R^2w^2}{\alpha^{\prime}(1-v^2)N^2}
+ \frac{\rho^2}{\pi^2\alpha^{\prime}}
\sin^2 \Big[ \frac{\sqrt{1-v^2}(\alpha -\beta)}{2}\pi \alpha^{\prime}E+
\pi \tilde q R\frac{w}{N} + \frac{\pi m}{N}\Big]
 + \hat N.\label{boost D0 open H}
\end{equation}
The multiplicity of the open string channel is the same as in the previous
case (\ref{trace}). Note that the Lorentz contraction factor
$\sqrt{1-v^2}$ does not appear in the open string 1-loop trace (\ref{trace})
because the open string energy $E$ is of the original coordinates picture.\\

If we take the T-duality (\ref{tdual}), a boosted D0-brane is 
transformed into 
a D1-brane with electric flux $f$. The boundary condition is written 
as follows
\ba
[\prt_{\tau}T+if\prt_{\sigma}Y]|_{\tau=0}|B\rangle=0,\ \ \
[\de_{\tau} Y+if\de_{\sigma} T]|_{\tau=0}|B\rangle=0.
\label{D1 condition }
\ea
The boundary state analysis shows that the condition of moving 
away from the origin is given by
\ba
\Big[q_++(1+f)\f{(\beta-\ap)}{2}\Big]R\in {\bf{Q}}. \label{d1m}
\ea
Again we can
choose the flux $f$ such that the D1-brane can move around.

Finally, we obtain the open string Hamiltonian of D1-brane
\begin{equation}
H_o = -\alpha^{\prime}E^2 +
\frac{\alpha^{\prime}n^2}{(1-f^2)R^2N^2}
+ \frac{\rho^2}{\pi^2\alpha^{\prime}}
\sin^2 \Big[ \frac{\sqrt{1-f^2}(\alpha -\beta)}{2}\pi \alpha^{\prime}E+
\pi \frac{(\alpha+\beta)}{2R}\alpha^{\prime}\frac{n}{N} 
+ \frac{\pi m}{N}\Big]
 + \hat N,\label{boost D1 open H}
\end{equation}
where we defined $R[q_+ +\frac{(1+f)}{2}(\beta-\alpha)]=\frac{k}{N}$.

\subsection{Flux Stabilizations of D-branes}
\hspace{5mm}
Up to now we have considered D-branes in the free field theory. 
Even though this viewpoint is the 
most convenient for calculations, it is better
for the study of geometric structure of these D-branes to examine
in terms of the original coordinate $(\rho,\varphi,t,y)$. This analysis 
can be done by extending the method in \cite{TU3}, where D2-branes in 
Melvin background were shown to be stabilized by the magnetic
flux $B_{y\vp}+F_{y\vp}$. In our curved spacetime model we will see 
the electric flux also
plays an important role. This explains the novel fact that 
we can move the D2-branes away from the origin for any value of 
$\ap,\beta$ 
by adding the electric flux, which we have already found in the boundary
state analysis.

The boundary condition 
in terms of the original coordinate can be obtained by the transformation
of fields under the T-duality (see also \cite{TU2})
\ba
&&\de\tvp=\f{\rho^2}{1+\ap\beta\rho^2}\left[-\de\vp-(q_++\beta)\de y+
(q_-+\beta)\de t\right],\no
&&\db\tvp=\f{\rho^2}{1+\ap\beta\rho^2}\left[\db\vp+(q_+-\ap)\db y
-(q_-+\ap)\db t\right],\no
&&\de\tvp=-\rho^2\de\vp'',\ \ \ 
\db\tvp=\rho^2\db\vp''. \label{CH}
\ea

The open string boundary condition of a D0-brane (including boosted ones)
is given by
\ba
\de_1(T+vY)=0,\ \ \ \de_2(vT+Y)=0, \ \ \de_2 \vp''=0,\label{D1 condition}
\ea
where the coordinates $(\sigma_1,\sigma_2)$ represent the
world sheet of the open string channel.
Then by applying (\ref{CH}) and comparing the mixed boundary condition
\ba
G_{\mu\nu}\de_1 X^\nu+i(B_{\mu\nu}+F_{\mu\nu})\de_2X^\nu=0,
\ea
we can identify the D-brane as a D2-D0 bound state wrapped on the two
dimensional torus in the directions of $y$ and $\varphi$. In other words,
this corresponds to D2-branes with the
following
the electric and magnetic flux 
\ba
F_{\hat{\vp}t}=\f{2v}{\ap+\beta+v(\beta-\ap)}\equiv vh,\ \ 
F_{\hat{\vp}y}=\f{2}{\ap+\beta+v(\beta-\ap)}= h,\label{dflux}
\ea
where we define the coordinate $\hat{\vp}=\vp+(q_++\f{\beta-\ap}{2})y
-(q_-+\f{\ap+\beta}{2})t$. 

The important point is that the flux quantization
$\frac{1}{4\pi^2 \alpha^{\prime}}
\int_{\bf{T^2}} \mbox{Tr}\, F \in {\bf Z}$ can
be easily satisfied by choosing the value of 
velocity $v$ suitably if $\ap\neq\beta$. 
This makes a striking contrast with the D-branes in 
Melvin background $(\beta=\ap)$. This fact is consistent with the boundary 
state analysis since the twisting parameter given by 
$\gamma=\al\f{\ap+\beta+v(\beta-\ap)}{2}\f{n}{R}$ should always be rational
numbers. The quantization condition is given by
\begin{equation}
h=\frac{\alpha^{\prime}N}{kR}, \label{h}
\end{equation}
where $k$ and $N$ represent the number of D2-branes and D0-branes,
respectively.

We can also check that the energy of D2-brane wrapped on the torus by 
using DBI-action is independent of $\rho$ and reproduces the mass of 
D0-branes as we show below.
 
The matrix $G+B+F$ in the $(t,y,\hat \varphi)$ coordinate system is
given by
\begin{equation}
(G+B+F)_{\mu\nu} =\frac{1}{H}\left(\begin{array}{ccc}
-1 -\frac{(\alpha +\beta)^2}{4}\rho^2&\frac{(\beta^2-\alpha^2)}{4}\rho^2&
\frac{(\alpha -\beta)}{2}\rho^2 -vhH\\
\frac{(\beta^2-\alpha^2)}{4}\rho^2&1-\frac{(\alpha -\beta)^2}{4}\rho^2&
\frac{\alpha +\beta}{2}\rho^2-hH\\
-\frac{(\alpha -\beta)}{2}\rho^2 +vhH&-\frac{\alpha +\beta}{2}\rho^2+hH&
\rho^2
\end{array}\right) \label{G+B+F},
\end{equation}
where we defined $H \equiv 1+\alpha\beta \rho^2$.
By applying (\ref{h}),
we can calculate the DBI action of the D2-D0 bound state
\ba
S&=& \int \d t \int \d y \int \d \hat \varphi \frac{e^{-\phi}}{4\pi^2
(\alpha^{\prime})^{\frac{3}{2}}}\mbox{Tr}\sqrt{-\det (G+B+F)}\no
&=& \frac{V_0kRe^{-\phi_0}}{(\alpha^{\prime})^{\frac{3}{2}}}\,
\sqrt{-H\det(G+B+F)},
\ea
where we used the relations,
$\phi = \phi_0 - \frac{1}{2}\ln H$.
Now we can calculate from eq.(\ref{G+B+F})
\begin{equation}
-H\det(G+B+F)=\rho^2\Big[\Big( 
\frac{\alpha +\beta +v(\beta -\alpha)}{2}\Big)h-1\Big]^2 +(1-v^2)h^2.
\label{acd}
\end{equation}
Then we find that the dependence of (\ref{acd}) on $\rho$ 
vanishes only when
the specific value (\ref{dflux}) of the flux is satisfied. 
Finally we obtain
 by using the relation (\ref{h})
\begin{equation}
S=\sqrt{1-v^2}V_0 \times NT_{D0}, \quad
(T_{D0}\equiv e^{-\phi_0}\alpha^{\prime -\frac{1}{2}}).
\end{equation}
This is the action of the bulk D0-brane
in the free field picture (including the
Lorentz contraction factor).\\

Similarly, the boundary condition of D1-branes 
in the free field theory with
the electric flux $f$, which we can obtain from
eq.(\ref{D1 condition}) by taking the T-duality 
of a D0-brane, is given by
\ba
\de_1 T-if\de_2 Y=0,\ \ \
\de_1 Y-if\de_2 T=0,\ \ \
\de_2\vp''=0. 
\label{D1 condition open}
\ea

In the same method as before we can determine its interpretation in the 
curved spacetime (\ref{KK}). The result is a D1-brane wrapped on the spiral
string 
\ba
\hat{\vp}+\f{f}{2}(\beta-\ap)y-\f{f}{2}(\ap+\beta)t=\mbox{const},\label{geo}
\ea
with the electric flux $F_{yt}=-f$.
This will be understood as the geodesic surface in the spacetime.
>From (\ref{geo}) we can explain the previous condition (\ref{d1m})
of moving  away from the origin if we remember
the periodicity $y\sim y+2\pi R$ and
$\vp\sim\vp+2\pi$ as in \cite{DuMo,TU2}.

\subsection{D-brane Instanton}
\hspace{5mm}
In our model the time direction is also curved as well as space directions
and thus it will be interesting to examine D-brane instantons\footnote{
Recently, D-brane instantons were reinterpreted from the viewpoint of
tachyon condensation \cite{GuSt2}.}, which have the
Dirichlet boundary condition in the time direction
\ba
\qquad \partial_{\sigma} T|_{\tau=0}
|B \rangle=0 : \quad
(\beta -\alpha)\hat J_0|B \rangle=0 .\label{dt}
\ea
The condition (\ref{dt}) shows that $\hat{J}$ should be zero
unless $\ap=\beta$ (Melvin model). 
Then (below we assume $\ap\neq\beta$) we can construct a D0-brane 
(and D1-brane) only at 
the origin and they cannot move. 
A D2-brane (and D3-brane) can also exist. These results would be useful 
when we consider the non-perturbative effect by D-instantons; they show that
we can neglect the instanton effect of D0 and D1-branes if we concentrate on physics at the 
large value of $\rho$.

\section{Penrose Limit in NSNS Background}
\setcounter{equation}{0}
\hspace{5mm}
In this section we would like to discuss the Penrose Limit \cite{Pe} of
superstring backgrounds with NSNS flux and to
apply our previous results
on D-branes to these limits.
The Penrose limit gives us the notion of a Lorentzian version of 
tangent plane. Thus this is a useful tool of approximating a
curved spacetime. This limit has recently been applied to the 
background $AdS_p\times S^q$ with RR-flux 
and found to be the maximally supersymmetric pp-wave background
\cite{BF,BeMaNa}.
This background is exactly solvable in the light-cone Green-Schwarz string
theory \cite{Me,BeMaNa} (see also \cite{RuTs,BMNstring}) and
the duality between
gauge and gravity theory has been checked even 
for stringy excitations \cite{BeMaNa} (for general
discussions on the holographic
relation see also \cite{holography}).
Here we mainly concentrate on
the background with NSNS flux.

The most simplest example is the Penrose limit of NS5-branes. This
was analyzed in \cite{GoOo} and it was
found that the limit is equivalent to the product of Nappi-Witten model 
(NW model) 
\cite{NaWi} defined by the action (\ref{NP}) 
and the six dimensional flat space ${\bf{R}}^6$. 
The half of the maximal
supersymmetries are preserved in this model, 
which is the same number as in the 
original NS5-branes.

It is interesting to apply this limit to 
the near horizon limit of F1-N5 system, that is,
$AdS_3\times S^3$ \cite{BeMaNa,RuTs}. The result
is given by the six dimensional generalization of NW model.
 Since we have the half of the maximal
supersymmetries in all of these kinds of NW model, this again 
reproduces
the original number of supersymmetries (sixteen) in $AdS_3\times S^3$.

\subsection{Supersymmetry Enhancement in the Penrose Limit of\\
Spacetime Coset Model}
\hspace{5mm}
One of the interesting phenomena observed in the discussion on the
Penrose limit of RR-flux background is the enhancement of 
supersymmetry\footnote{For the discussions of supersymmetry of pp-waves 
in a more general context and its connection with the worldseet
supersymmetry of the massive string action see also \cite{Cv}.} 
\cite{ItKlMu,GoOo,ZaSo}. This phenomenon was found in the example
of $AdS_5\times T^{1,1}$, which preserves $1/4$ of the maximal
supersymmetries. The manifold $T^{1,1}$ is defined to be a coset
$SU(2)\times SU(2)/U(1)$. If we take the Penrose limit in this
model, it is enhanced
maximally \cite{ItKlMu,GoOo,ZaSo}. In the appendix \ref{W} 
we show that 
such an enhancement also occurs in another supersymmetric coset system 
$W_{4,2}\times S^5$ 
considered in \cite{BrStVl}, where $W_{4,2}$ is a Lorentzian coset of
$SL(2,{\bf R})\times SL(2,{\bf R})/U(1)$. 
On the other hand, if we consider
orbifolds of the pp-waves, then the supersymmetry is generically\footnote{
Recently, the example of supersymmetry enhancement in orbifolded pp-waves
was also interpreted holographically \cite{pporbifolde}.}
reduced and can be interpreted in terms of the dual quiver gauge theory
\cite{ItKlMu,pporbifold}.

Now we are interested in possibilities of 
supersymmetry enhancement in
the background with only NSNS flux. As a particular example
we would like to examine 
the Penrose limit of the model constructed in \cite{ZaTs}.
The spacetime in this model is described by the coset SCFT
\ba
\f{SL(2,{\bf R})_{k_1+2}\times SL(2,{\bf R})_{k_2+2}}{U(1)}
\times \f{SU(2)_{k_1-2}\times SU(2)_{k_2-2}}{U(1)},\label{slsu}
\ea
where $k_1$ and $k_2$ are the level of the WZW model and we can see
the total central charge of the model agrees with the correct value.
We denote the first (non-compact) manifold by $W$ and 
the second by $T$. It is known that this model is non-supersymmetric 
\cite{ZaTs}.

The explicit metric is given by ($k\equiv k_1$ and $Q^2 \equiv k_2/k_1$)
\ba
ds^2&=&k[(d\theta_1)^2+\sin^2\theta_1(d\phi_1)^2]
+kQ^2[(d\theta_2)^2+\sin^2\theta_2(d\phi_2)^2]\no
&&+k(d\psi+\cos\theta_1d\phi_1+Q\cos\theta_2d\phi_2)^2\no
&&+k[(dr_1)^2+\sinh^2r_1(d\tp_1)^2]+kQ^2[(dr_2)^2+\sinh^2r_2(d\tp_2)^2]\no
&&-k(dt+\cosh r_1d\tp +Q\cosh r_2d\tp_2)^2.\label{slsu2}
\ea
Then the metric of $T$ (the former part) 
shows the coset space is $T^{1,Q}$ type 
even though it is not Einstein metric. It satisfies the equation of motion
with the non-trivial B-field
\ba
B=k\cos\theta_1d\phi_1\we d\psi+kQ\cos\theta_1\cos\theta_2 d\phi_1\we d\phi_2
-kQ\cos\theta_2d\phi_2\we d\psi.
\ea
In order to discuss its Penrose limit,
we would like to take the large volume limit $k\to \infty$.
Let us take the following limit
\ba
&&x^{+}=\f{1}{2}\left(t+\tp_1+Q\tp_2+\psi+\phi_1+Q\phi_2\right),\ \ \
x^-=\f{k}{2}\left(t+\tp_1+Q\tp_2-(\psi+\phi_1+Q\phi_2)\right),\no
&&\theta_1=\xi_1/{\s{k}},\ \ \theta_2=\xi_2/{\s{k}Q},\ \ 
r_1=\rho_1/{\s{k}},\ \ r_2=\rho_2/{\s{k}Q}.
\ea
Finally we obtain the pp-wave metric
\ba
ds^2&=&-4dx^+dx^- 
-(\xi_1^2d\phi_1+\f{\xi_2^2}{Q}d\phi_2+\rho_1^2d\tp+\f{\rho^2_2}{Q}d\tp_2)
dx^+ \no
&&+\sum_{i=1}^{2}(d\xi_i^2+\xi_i^2d\phi_i^2)
+\sum_{i=1}^{2}(d\rho_i^2+\rho_i^2d\tp_i^2).\label{mp}
\ea
and the $B$-field
\ba
B&=&-(\f{\xi_1^2}{2}d\phi_1+\f{\xi_2^2}{2Q}d\phi_2)\we dx^+ 
-(\f{\rho_1^2}{2}d\tp_1+\f{\rho_2^2}{2Q}d\tp_2)\we dx^+ +(trivial\ part).
\label{bp}
\ea

On the other hand, the ten dimensional NW model\footnote{
For higher dimensional NW models see \cite{Ke}.} is defined by 
the action 
\ba
S=\f{1}{\pi\al}\int(d\sigma)^2[\de u\db v
+\sum_{i=1}^{4}(\beta_i\ \rho_i^2\ \de u\db \phi_i
+\de\rho_i\db\rho_i+\rho_i^2\de\phi_i\db\phi_i)].\label{gnp}
\ea
Then our results (\ref{mp}) and (\ref{bp}) show that the Penrose limit of 
(\ref{slsu}) is equivalent to the generalized NW model (\ref{gnp})
with the specific parameters $\beta_1=\beta_3=-1,\ \beta_2=\beta_4=-1/Q$.

The spectrum can be computed as follows
\ba
E^2-\vec{p}^2=\f{2}{\al}(\hat{N}_L+\hat{N}_R)+p_y^2-2(p_y+E)(\sum_{i=1}^4
\beta_i\hat{J}_{iR}).
\ea

In this limit we can conclude the half of maximal supersymmetries are enhanced
in spite of starting with the non-supersymmetric model. This implies that 
there are supersymmetric sectors in the whole non-supersymmetric holographic
dual theory.

\subsection{D-branes in Nappi-Witten Background}
\hspace{5mm}
Next let us apply our previous results of D-branes to the Penrose 
limit\footnote{
For recent discussions on D-branes in pp-wave (RR) background see 
\cite{BMNopen}.}of the model (\ref{S1}), namely NW model. 
Since the supersymmetry of the bulk sector 
is restored in this limit, here we are interested in the supersymmetry of
D-branes (for the analysis of D-branes in NW model from the viewpoint of 
current algebra see \cite{StTs}).
The NW model, as we have seen in section \ref{sec:PPNW}, can be obtained
by setting $R\to\infty$ and $q_\pm=0,\ap=0$. Because in this limit there
are no winding modes in the boundary condition (\ref{boundary condition 
in Y frame}), we must always set $\hat{J}=0$ for a D0-brane. 
Thus a D0-brane can 
only exist at the origin\footnote{
Even if we consider the boosted D0-branes as before in order to move around, 
we must set $v=-1$ and 
thus they become singular.}. An important result is that the supersymmetry
seems to be broken on this brane as Bose-Fermi degeneracy can
not be seen in the vacuum 
amplitude (\ref{fractional D0 1}). 
The corresponding open string Hamiltonian is given by
\ba
H_o=-\al E^2+\f{\al}{4}\beta^2\hat{J}^2+\hat{N}.
\ea
Note that we can see that there are no tachyons on D-branes.
On the other hand, the D1-brane in this limit 
has the similar property to
a D1-brane in flat space. For example, it is easy to see that 
its vacuum amplitude is the same as in the flat space and thus does vanish.
It can also move away from the origin.

\section{Conclusions and Discussions}
\hspace{5mm}
In this paper we investigated the
superstring version of the exactly solvable model \cite{3RuTs}, which 
describes a curved spacetime with four parameters. In particular, 
we constructed the boundary 
states of D-branes in this model and calculated their open string spectra. 
Generally, in this model
supersymmetries are completely broken and closed string tachyons
appear.
For a particular region of the parameters we have spacetime singularities.
However, we show that the open string 
spectrum does not become singular at high energy
for any value
of parameters due to $\al$ corrections. This seems to 
suggest that the phenomenon is due to a stringy resolution
of spacetime singularity and we may allow such a singularity. 
In other words, the free field representation of the background
may give a kind of analytic continuation of the singular spacetime.
One of the well-studied example of space singularities smoothed in
string theory is the orbifold 
singularity (see e.g. \cite{As}). Our example may be another type of 
resolution of spacetime singularities.
We would like to leave the detailed interpretation
of this as a future problem. 

Though our analysis of D-branes 
can be seen as a generalization of the previous analysis on 
D-branes in Melvin background \cite{DuMo,TU2,TU3}, we found a crucial 
difference. We show that we can always construct bulk (or movable) 
D0-branes 
in the analysis of the free field representation
except the parameter region where it is equivalent to the 
Melvin model \cite{SM}. This result is consistent with the geometrical
interpretation of a bulk D0-brane as a D2-D0 bound state wrapped on a torus 
with an electric and magnetic flux.
This phenomenon implies that the geometry is smooth 
as the brane can probe the whole spacetime. 

Furthermore, our model includes the Nappi-Witten model
as the Penrose limit.
We examined D-branes in this model by using the general results.
The Nappi-Witten like models can appear in many examples of Penrose limits
of backgrounds with NSNS $B$ flux. 
Especially, we showed the non-supersymmetric spacetime coset model
becomes the higher dimensional Nappi-Witten model and thus supersymmetric.
Such an enhancement of supersymmetry can also be seen in the Penrose limit of 
$1/4$ BPS
spacetime coset model with RR-flux and this may be a clue to investigating
its holographic dual gauge theory.

\vskip6mm
\noindent
{\bf Acknowledgments}

\vskip2mm
We would like to thank T. Eguchi for valuable advice and encouragements.
We are grateful to Y. Hikida, K. Sakai, Y. Sugawara, T. Uesugi 
and S. Yamaguchi
for stimulating discussions.
The research of T.T
was supported in part by JSPS Research Fellowships for Young
Scientists.

\appendix

\section{Fractional Branes}
\label{Ap:frac}
\setcounter{equation}{0}
\hspace{5mm}
Here we show that the (boosted) fractional D0-branes satisfy
the Cardy's condition. For detailed convention of boundary
states see \cite{TU2}.

\subsubsection*{Fractional D0-brane}
\hspace{5mm}
First, let us consider a fractional D0-brane.
Its boundary state is
\begin{equation}
|B\rangle = \sum_{n=-\infty}^{\infty}f_{\gamma}|n \rishi \otimes
|B,\gamma \rishi_{NSNS,RR},
\end{equation}
where the total boundary state $|B\rangle$ 
consists of the summation of boundary state $|n \rishi \otimes
|B,\gamma \rishi_{NSNS,RR}$
with fixed momentum $n$ 
in the ${\bf S}^1$ direction.
Its vacuum amplitude is given by
\ba
&&{\cal A} = \frac{\alpha^{\prime}V_0}{8\pi R} \sum_{\gamma \in {\bf Z}}
|f_{\gamma}|^2\int \d s \, (2\pi \alpha^{\prime} s)^{-4}
\exp \Big(-\frac{s \alpha^{\prime}n^2}{2R^2} \Big)\no
&&\qquad \times
\eta(\tau)^{-12}[\theta_3(0|\tau)^4-(-1)^{\gamma}\theta_4(0|\tau)^4
-\theta_2(0|\tau)^4]\no
&&\quad + \frac{i\alpha^{\prime}V_0}{8\pi R}\sum_{\gamma \not \in {\bf Z}}
(-1)^{[\gamma]}|f_{\gamma}|^2 \int \d s \, (2\pi \alpha^{\prime} s)^{-3}
\exp \Big(-\frac{s \alpha^{\prime}n^2}{2R^2}\Big) \eta(\tau)^{-9}
\theta_1(\nu|\tau)^{-1}\no
&&\qquad \times [\theta_3(0|\tau)^3\theta_3(\nu|\tau)
-\theta_4(0|\tau)^3\theta_4(\nu|\tau)
-\theta_2(0|\tau)^3\theta_2(\nu|\tau)],
\label{fractional D0 1}
\ea
where $\tau =\frac{is}{\pi},\nu=\gamma \tau$.

On the other hand, the open string 1-loop amplitude is
\begin{equation}
Z_o = 2\times \int_{0}^{\infty} \frac{\d t}{2t} \, \mbox{Tr}_{NS-R}
\Big[\frac{1+(-1)^F}{2}q^{H_o}\Big] \qquad , \quad q \equiv e^{-2\pi t},
\label{ZO fractional D0}
\end{equation}
where we have defined $\mbox{Tr}_{NS-R}=\mbox{Tr}_{NS}-\mbox{Tr}_{R}$;
the operator $H_o$ denotes the open string Hamiltonian
\begin{equation}
H_o=-\alpha^{\prime}E^2 + \alpha^{\prime}
\Big( \frac{Rw}{\alpha^{\prime}} -\frac{\alpha+\beta}{2}
\hat J\Big)^2+\hat N.
\end{equation}

By requiring the equality
between eq.(\ref{fractional D0 1}) and (\ref{ZO fractional D0}),
we obtain
\begin{equation}
\gamma \in {\bf Z}: \quad f_{\gamma} = \frac{T_0}{2} \quad , \quad
\gamma \not \in {\bf Z}: \quad f_{\gamma}=\frac{1}{\sqrt{2}}\Big(
\frac{|\sin \pi \gamma |}{2\pi^2\alpha^{\prime}}\Big)^{\frac{1}{2}}T_0,
\end{equation}
where we have defined $T_0 = \sqrt{\pi}(2\pi\sqrt{\alpha^{\prime}})^3$.
\subsubsection*{Bulk D0-brane at $\rho=0$}
\hspace{5mm}
If the parameter $\frac{(\alpha+\beta)}{2R}\alpha^{\prime}$ is rational
($\equiv
\frac{k}{N}$), there is a bulk D0-brane.
At first, by using $U(1)$ phase in the coefficient $f_{\gamma}$,
we can write a fractional brane at $Y=y_0$ as
\begin{equation}
|B ,y_0 \rangle = \frac{1}{2\pi R}\sum_n f_{\gamma}e^{i\frac{n}{R}(\hat
y-y_0)}
\Big[|\gamma,+ \rangle_{NS} - (-1)^{[\gamma]}|\gamma ,- \rangle_{NS}
+|\gamma , + \rangle_R +(-1)^{[\gamma]}|\gamma ,- \rangle_{R}\Big].
\end{equation}
If we set $N$ fractional D0-branes on the condition
\begin{equation}
|B \rangle_{bulk} \equiv \sum_{a=0}^{N-1}|B,y_a \rangle \qquad , \quad
y_a \equiv \tilde y_0 + \frac{2\pi R}{N}a \quad ,
\quad 0 \leq \tilde y_0 < \frac{2\pi R}{N},
\end{equation}
only the sectors of $n=N\ell \, (\gamma = k\ell \in {\bf Z})$
survive, then $|B \rangle_{bulk}$ describes a bulk D0-brane at
$\rho=0$.
If we set $\tilde y_0=0$ for simplicity, the explicit form of
$|B \rangle_{bulk}$ is
\begin{equation}
|B \rangle_{bulk} = \frac{N}{2(2\pi R)}\sum_{\ell}\frac{T_0}{2}
e^{i\frac{N}{R}\ell \hat y}\Big[
|0 ,+ \rangle_{NS} -(-1)^{k\ell}|0,- \rangle_{NS}
+|0,+ \rangle_R +(-1)^{k\ell}|0,- \rangle_R \Big].
\label{boundary state of bulk D0 at origin}
\end{equation}
The amplitude between two $|B \rangle_{bulk}$ is
\ba
&&{\cal A}_{bulk} =\frac{\alpha^{\prime}V_0}{8\pi R}
\sum_{\ell=-\infty}^{\infty}
\Big(\frac{NT_0}{2}\Big)^2
\int \d s \, (2\pi \alpha^{\prime} s)^{-4}
\exp \Big(-\frac{s \alpha^{\prime}(N\ell)^2}{2R^2} \Big)\no
&&\qquad \qquad \times
\eta(\tau)^{-12}[\theta_3(0|\tau)^4-(-1)^{k\ell}\theta_4(0|\tau)^4
-\theta_2(0|\tau)^4].
\label{bulk D0 amplitude at rho 0}
\ea
Therefore by performing the modular transformation 
of eq.(\ref{bulk D0 amplitude at rho
0}),
we obtain
\begin{equation}
Z_o = 2N\times \int_{0}^{\infty} \frac{\d t}{2t} \, \mbox{Tr}_{NS-R}
\Big[\frac{1+(-1)^F}{2}q^{H_o}\Big],
\label{ZO bulk D0}
\end{equation}
where $H_o$ is
\begin{equation}
H_o = -\alpha^{\prime}E^2 +\frac{R^2}{\alpha^{\prime}N^2}
(w -k\hat J)^2 + \hat N.
\label{open Hamiltonian at origin}
\end{equation}
\subsubsection*{Boosted Fractional D0-brane}
\hspace{5mm}
Then we consider the boosted fractional D0-branes. First, we must
add the Lorentz contraction factor $\sqrt{1-v^2}$
to the vacuum amplitude ${\cal A}$. This factor is absorbed
when we perform the Poisson resummation formula with respect to $n$
$$
\sum_{n}\sqrt{(1-v^2)}
\exp \Big[-\frac{\alpha^{\prime}(1-v^2)}{2R^2t}n^2 + 2\pi i 
\gamma_{v0}n\hat J \Big]=\sqrt{\frac{2R^2t}{\alpha^{\prime}}}
\sum_{w}\exp \Big[-\frac{2\pi R^2t}{\alpha^{\prime}(1-v^2)}
(w-\gamma_{v0}\hat J)^2 \Big].
$$
Then we find that the boosted fractional D0-branes satisfy
the Cardy's condition and the open string Hamiltonian is
\begin{equation}
H_o=-\alpha^{\prime}E^2+ \frac{R^2}{\alpha^{\prime}(1-v^2)}
(w-\gamma_{v0}\hat J)^2 +\hat N.
\end{equation}
\subsubsection*{Boosted Bulk D0-brane at $\rho=0$}
\hspace{5mm}
Finally, when $\gamma_{v0}=\frac{k}{N}$, 
the boosted bulk D0-brane at $\rho=0$ consists of
$N$ boosted fractional
D0-branes whose positions are $Y=y_0,
y_0+\frac{2\pi R}{N},\cdots$. Its open string
Hamiltonian is given by
\begin{equation}
H_o = -\al E^2+ \frac{R^2}{\alpha^{\prime}N^2(1-v^2)}
(w-k\hat J)^2 +\hat N.
\end{equation}
The multiplicity of the open string channel is the same as before.

\section{Bulk D0-brane in Terms of Bessel Functions}
\label{Ap:Be}
\setcounter{equation}{0}
\hspace{5mm}
Here we show the detailed calculations 
of boundary state with $\rho\neq 0$
by using the Bessel functions. Even though the 
analysis in section 3.2 can be done without
the Bessel function representation, 
the geometrical interpretation will be more transparent by using them.
In particular, this would be helpful if one computes the couplings of 
D-branes with closed string.

The only non-trivial part is
the bosonic zero modes $(x_1,x_2)$ (or $(k_1,k_2)$)
of two dimensional free fields $(X_1,X_2)$, which represent two dimensional
plane. It is equivalent to 
quantum mechanics of a free particle and below we will use the 
polar coordinates $(\rho,\vp)$.
The Hamiltonian is
\begin{equation}
2H =-\frac{\alpha^{\prime}}{2}\Big(\frac{\prt^2}{\prt x_1^2}+
\frac{\prt^2}{\prt x_2^2}\Big)
=-\frac{\alpha^{\prime}}{2}\Big[
\frac{1}{\rho^2}\frac{\prt^2}{\prt \varphi^2}+\frac{1}{\rho}
\frac{\prt}{\prt \rho}\Big(\rho \frac{\prt}{\prt \rho}\Big)\Big].
\label{closed H in polar coordinates}
\end{equation}
Its eigen functions and eigen values are\footnote{Here we impose the
the smooth boundary condition at the origin $\psi(0 ,\varphi)=$finite
as usual and thus we do not consider the independent solutions of 
$N_m(k\rho)$ (Neumann function).}
\begin{equation}
\psi(\rho ,\varphi) =J_m(k\rho)\, e^{im\varphi}
\qquad , \quad 2H \psi (\rho , \varphi) =
\frac{\alpha^{\prime}}{2}k^2 \psi (\rho , \varphi),
\end{equation}
where $k$ is the magnitude of the momentum which takes non-negative value
and
$m$ is the angular momentum which takes integer ($=\hat{J}$) values.
$J_m(z)$ is Bessel function defined as
\begin{equation}
J_m(z) \equiv
\Big(\frac{z}{2}\Big)^m\sum_{n=0}^{\infty}\frac{(-1)^n(z/2)^{2n}}
{n! (m+n)!} \quad , \quad J_{-m}(z)\equiv(-1)^mJ_m(z)
\quad \mbox{for}\quad m \geq 0.
\end{equation}
The generating function of Bessel functions is given by
\begin{equation}
e^{iz\sin \theta}=\sum_{m=-\infty}^{\infty}J_m(z)e^{i m \theta}.
\label{generating function of Bessel}
\end{equation}
Then we can expand a plane wave in terms of Bessel functions
\begin{equation}
e^{i{\bf k}\cdot ({\bf x}-{\bf x_0})}
=\sum_{m , n=-\infty}^{\infty}
J_m(k\rho)J_n(k\rho_0)(-1)^n e^{im\varphi}e^{in\varphi_0}
e^{i(m+n)\theta},
\label{plane wave in Bessel units}
\end{equation}
where
\begin{equation}
{\bf x} \equiv (\rho \cos \varphi , \rho \sin \varphi), \quad
{\bf x_0} \equiv (\rho_0 \cos \varphi_0 , \rho_0 \sin \varphi_0),
\quad {\bf k} \equiv (k\cos \theta, k\sin \theta).
\end{equation}
By using eq.(\ref{plane wave in Bessel units}), we obtain
the expansion of a delta function
\begin{equation}
\delta^2({\bf x}-{\bf x_0})
=\frac{1}{2\pi}\sum_{m=-\infty}^{\infty} \int_{0}^{\infty}\d k \, k
J_m(k\rho)J_m(k\rho_0)e^{im(\varphi -\varphi_0)}.
\end{equation}
Finally, let us consider the superposition of $N$ delta functions at
\begin{equation}
{\bf x}_{0a} =(\rho_0 , \varphi_0 + \frac{2\pi}{N}a)
\qquad , \quad a=0,\cdots , N-1.
\end{equation}

Then, the wave function of a bulk D0-brane at $\rho_0 \neq 0$ is written as
follows
\begin{equation}
\psi_{bulk}({\bf x})= \frac{1}{N}\sum_{a=0}^{N-1}\delta^2({\bf x}-{\bf
x}_{0a})
=\frac{1}{2\pi}\sum_{j=-\infty}^{\infty}
\int_0^{\infty}\d k\, k
J_{Nj}(k\rho)J_{Nj}(k\rho_0)e^{-iNj(\varphi-\varphi_0)}.
\label{bulk D0 Bessel}
\end{equation}
Notice that only $m \equiv 0\ (\mbox{mod} N)$ states
survive in eq.(\ref{bulk D0 Bessel}).
This is consistent with the boundary condition
(\ref{TNYD boundary condition again}).
Also if we set $\rho_0=0$, only $m=0$ states survive
because $J_m(0)=0$ for $m\neq 0$. This is also consistent with
the fact that $\hat J_0=0$ when a D0-brane is at the origin.
\subsubsection*{Vacuum Amplitude}
Next we calculate the vacuum amplitude of the bulk D0-brane at
$\rho_0$
\begin{eqnarray}
{\cal A}_X &\equiv& \int \d^2 x \, \psi_{bulk}^{\ast}({\bf x})
e^{-2(H+\Delta H)s}
\psi_{bulk}({\bf x})\no
&=&\frac{1}{2\pi}\int_{0}^{\infty} \d k \, k
\sum_{j=-\infty}^{\infty} J^2_{Nj}(k\rho_0)
e^{-\frac{\alpha^{\prime}}{2}k^2s}
e^{\frac{(\alpha-\beta)^2N^2}{8}\alpha^{\prime}sj^2},
\label{amplitude in appendix1}
\end{eqnarray}
where $2\Delta H = -\frac{(\alpha-\beta)^2}{8}\alpha^{\prime}\hat J_0^2$.
This is equivalent to the calculation
in the section \ref{sec:bulk D0 at non zero
rho} with $\tilde q=0$.
By using the formula
\begin{equation}
J_n^2(z)=\frac{1}{2\pi}\int_{-\pi}^{\pi}\d \theta
\, e^{2in\theta}\, J_0(2z\sin \theta),
\end{equation}
we can rewrite (\ref{amplitude in appendix1}) as
\begin{equation}
{\cal A}_X =
\frac{1}{(2\pi)^2}\int_{0}^{\infty} \d k \, k
e^{-\frac{\alpha^{\prime}}{2}k^2s}
\int_{-\pi}^{\pi}\d \theta \, J_0(2k\rho_0 \sin \theta)
\sum_{j=-\infty}^{\infty}\exp \Big[
\frac{\pi(\alpha -\beta)^2\alpha^{\prime}N^2}{8t}j^2+2\pi i
\frac{\theta N}{\pi}j\Big].
\label{amplitude in appendix2}
\end{equation}
Next, after performing the Poisson resummation with respect to $j$,
we integrate out $k$ by using the formula
\begin{equation}
\int_{0}^{\infty}\d x \, x e^{-a^2x^2}J_0(bx)
=\frac{e^{-b^2/4a^2}}{2a^2}.
\end{equation}
Then eq.(\ref{amplitude in appendix2}) changes to
\begin{equation}
{\cal A}_X = \frac{1}{2\pi N}\frac{i\sqrt{2\alpha^{\prime}t}}
{(\alpha -\beta)\alpha^{\prime}\pi}
\sum_{m=-\infty}^{\infty}\int_{-\pi}^{\pi}\d \theta \,
\frac{\exp \Big[ -\frac{2\rho_0^2\sin^2 \theta}{\alpha^{\prime}s}\Big]
}{\alpha^{\prime}s}
\exp \Big[ \frac{8\pi t}{(\alpha -\beta)^2\alpha^{\prime}}
\Big(\frac{\theta}{\pi}-\frac{m}{N}\Big)^2\Big].
\end{equation}
Since we can rewrite the integral of $\theta$ by employing
\begin{equation}
\sum_{m=-\infty}^{\infty}\int_{-\pi}^{\pi}\d \theta \,
f(\sin^2 \theta)g(\frac{\theta}{\pi}-\frac{m}{N})=
2\sum_{m=0}^{N-1}\int_{-\infty}^{\infty}\d \theta \,
f \big[ \sin^2 (\theta +\frac{m}{N}\pi)\big]
g(\frac{\theta}{\pi}),
\end{equation}
we obtain
\begin{equation}
{\cal A}_X = \frac{2}{2\pi N}\frac{i\sqrt{2\alpha^{\prime}t}}
{(\alpha -\beta)\alpha^{\prime 2}s\pi} \sum_{m=0}^{N-1}
 \int_{-\infty}^{\infty}\d \theta \,
\exp \Big[ -\frac{2\rho_0^2\sin^2 (\theta +\frac{\pi m}{N})}
{\alpha^{\prime}s}+ \frac{8\pi t}{(\alpha -\beta)^2\alpha^{\prime}}
\Big(\frac{\theta}{\pi}\Big)^2\Big].\label{amplitude in appendix3}
\end{equation}

Finally, we can rewrite eq.(\ref{amplitude in appendix3}) by setting
$E=\frac{2\theta}{(\alpha -\beta)\alpha^{\prime}\pi}$
\ba
&&{\cal A}_X = \frac{i\sqrt{2\alpha^{\prime}t}}{2\pi \alpha^{\prime}s}
\frac{1}{N}\sum_{m=0}^{N-1}\int_{-\infty}^{\infty} \! \d E e^{-2\pi t
H_o}\no
&&H_o = -\alpha^{\prime}E^2 + \frac{\rho_0^2}{\pi^2\alpha^{\prime}}
\sin^2 \Big[ \frac{(\alpha -\beta)}{2}\pi \alpha^{\prime}E+\frac{\pi m}{N}
\Big].\label{appendix result}
\ea
These results are consistent with
eq.(\ref{open Hamiltonian at non zero rho}).

At last, if we take the limit
$\rho_0 \rightarrow 0$ in eq.(\ref{appendix result}),
${\cal A}_X$ approaches to $\frac{1}{2\pi \alpha^{\prime}s}$ which
is the value at $\rho_0=0$. This means that the limit
$\rho \rightarrow 0$ is smooth.

\section{Penrose Limit of Spacetime Coset with RR-flux}
\label{W}
\setcounter{equation}{0}
\hspace{5mm}
Here we examine the Penrose limit of the spacetime coset model 
$W_{4,2}\times S^5$, where $W_{4,2}$ is defined to be 
$SL(2,{\bf R})\times SL(2,{\bf R})/U(1)$.
This background was investigated in the context of
AdS/CFT correspondence in \cite{BrStVl}. There are eight supersymmetries
in this spacetime and the explicit metric is given by
\ba
(ds)^2&=&-\f{R^2}{9}(dt+\cosh y_1d\vp_1+\cosh y_2 d\vp_2)^2 
+\f{R^2}{6}\left(dy_1^2+\sinh^2 y_1 d\vp_1^2+\sinh^2 y_2 d\vp_2^2\right)
\no
&&+R^2(d\psi^2\cos^2\theta+d\theta^2+\sin^2\theta d\Omega_3^2).
\ea
We take the Penrose limit 
\ba
&&x^+=\f{1}{2}\left(\f{t+\vp_1+\vp_2}{3}+\psi\right),\ \  
x^-=\f{R^2}{2}\left(\f{t+\vp_1+\vp_2}{3}-\psi\right),
\ \ \rho=\f{\theta}{R},\ \no
&& y'_i=\f{\s{6}y_i}{R},\ \ \vp'=\vp-x^+,
\ea
and finally we obtain the maximally supersymmetric pp-wave background
\ba
ds^2=-4dx^+dx^- -(\rho^2+y_1^2+y_2^2)dx^{+2}+y_1^2d\vp_1^2
+y_2^2d\vp_2^2,
\ea
with the corresponding RR-flux. In this way we have observed that the 
$1/4$ supersymmetric background $W_{4,2}\times S^5$ will have enhanced 
supersymmetry in
the Penrose limit.
Interestingly, the final result turns out to be equivalent to the Penrose 
limit of
$1/4$ supersymmetric background $AdS_5 \times T^{1,1}$ 
discussed in \cite{ItKlMu,GoOo,ZaSo}. 
We hope this result
will be helpful for the analysis of holography in such a spacetime 
coset space.

\end{document}